\documentclass[epj-spec]{svjour}
\usepackage{graphicx}
\usepackage{amsmath,amsfonts,amssymb}
\usepackage{bm}

\newcommand{\be}{\begin{equation}}
\newcommand{\ee}{\end{equation}}
\newcommand{\la}{\label}
\newcommand{\p}{\partial}

\begin{document}
\title{A linear path toward synchronization}
\subtitle{Anomalous scaling in a new class of exactly solvable Kuramoto models}
\author{David C. Roberts and Razvan Teodorescu}
\institute{Center of Nonlinear Studies and Theoretical Division, Los Alamos National Laboratory}
\abstract{
Using a recently introduced linear reformulation of the Kuramoto model of self-synchronizing oscillator systems \cite{roberts}, we study a new class of analytically solvable oscillator systems defined by a particular coupling scheme.  We show that these systems have a logarithmic scaling law in the vicinity of the critical point, which may be seen as anomalous with respect to the usual power-law behavior exhibited by the standard Kuramoto model.
}
\maketitle
\section{Introduction}
Throughout the physical world there are examples of systems of coupled oscillators | Josephson junction arrays, swarms of flashing fireflies, cardiac pacemaker cells in a heart, etc. | that spontaneously synchronize under certain circumstances.  The Kuramoto model stands out as one of the few analytically soluble models that can describe the spontaneous synchronization phase transition of these systems \cite{strogatz1,acebron}.  For these reasons, it has become the paradigm in the field.  In addition, one may think of such self-synchronization as a type of nonequilibrium phase transition, which highlights the potential usefulness of the Kuramoto model as a way to study this latter phenomenon.  However, using the traditional approach, analytic solutions are largely restricted to systems where the number of oscillators approaches infinity (the continuum limit) and the oscillator populations are uniformly and globally coupled.

A recent paper introduced a linear version of this Kuramoto model \cite{roberts} and discussed the properties of the reformulation, in particular how it permitted solution of systems containing a finite number of oscillators.  In this paper, we work throughout in the continuum limit, where the linear reformulation of the Kuramoto model makes it possible to solve exactly a new set of coupled-oscillator systems.  We shall focus on these systems' interesting anomalous scaling properties about the phase transition.  However, before deriving and discussing these scaling properties, we will lay the foundation.  We will begin by demonstrating the mapping of the linear reformulation onto the Kuramoto model.  Next, we will derive the synchronization condition of the linear version.  We will then solve the synchronization order parameter and critical point exactly for a class of systems with a specific coupling scheme, which cannot be solved using the traditional approach to the Kuramoto model.  Finally we will outline the derivation of the anomalous scaling properties of the linear reformulation and discuss possible causes of this unusual behavior.

The linear model can be expressed as
\begin{equation}
\label{model}
\dot \psi(\omega,t)=(i \omega - \gamma) \psi(\omega,t) +\int_{-\infty}^{\infty} \Omega(\omega,\omega') g(\omega') \psi(\omega',t) d \omega'
\end{equation}
where $\Omega(\omega,\omega')$ describes the coupling between pairs of oscillators with characteristic frequencies $\omega$ and $\omega'$ respectively and the dot denotes a partial time derivative; the phases of the complex variable $\psi(\omega,t)$ correspond to those of the system's oscillators, the synchronization properties of which we will investigate; and $g(\omega')$ is the distribution of their characteristic frequencies. $\gamma$ is a parameter fixed according to the system parameters such that the amplitude of $\psi(\omega,t)$ goes to a steady state in the long-time limit.  This allows the linear model to be mapped onto the original Kuramoto model in the synchronized region, since with the nonlinear transformation $\psi(\omega)=R(\omega) e^{i \theta(\omega)}$ we can write the real and imaginary parts of eq. (\ref{model}) as
\begin{equation}
\dot R(\omega)=-\gamma R(\omega) +\int_{-\infty}^{\infty}\Omega(\omega,\omega') g(\omega') R(\omega') \cos[\theta(\omega')-\theta(\omega)] d \omega'
\end{equation}
\begin{equation}
\label{kurmap}
\dot \theta(\omega)=\omega+\int_{-\infty}^{\infty}\Omega(\omega,\omega') g( \omega') \frac{R(\omega')}{R(\omega)} \sin[\theta(\omega')-\theta(\omega)] d \omega',
\end{equation}
and, if $R(\omega)$ goes to a steady state in the long-time limit, eq. (\ref{kurmap}) is simply the Kuramoto model with a generalized coupling
\begin{equation} \la{coup}
K(\omega,\omega')=\Omega(\omega,\omega')  \frac{R(\omega')}{R(\omega)}.
\end{equation}
We discuss below how $\gamma$ is set such that this mapping can occur.

Having reformulated the Kuramoto model in terms of linear dynamics, we can proceed to apply to it the tools and techniques developed for solving linear problems. Indeed, the synchronization problem can be discussed in terms of the spectrum of the linear operator on the RHS of eq. (\ref{model}).  More 
precisely, let ${\cal K} (\omega, \omega') = \Omega(\omega, \omega') g(\omega') - i\omega \delta(\omega -\omega')$ and assume that the Fredholm integral equation 
\be \la{fred}
\int_{\mathbb{R}} d\omega' {\cal{K}}(\omega, \omega') \phi_\sigma(\omega') = \mu_\sigma \phi_\sigma(\omega), 
\quad \sigma \in \mathbb{Z}, \mathbb{R}
\ee
has a mixed, discrete-continuum spectrum $\{ \mu_n, \mu_\sigma \}$. Then a generic solution of (\ref{model}) is given by 
\be
\psi(\omega,t) = \sum_{n \in \mathbb{Z}} a_n \phi_n(\omega) e^{(\mu_n  -\gamma)t} + 
\int_{\sigma \in \mathbb{R}} b(\sigma) \phi_\sigma(\omega) e^{(\mu_\sigma  -\gamma)t} d \sigma, 
\ee
with coefficients $\{ a_n \}_{n \in \mathbb{Z}}, \{ b(\sigma)  \}_{\sigma \in \mathbb{R}}$ determined by initial conditions. 

To determine $\gamma$, we first solve for the spectrum $\{ \mu_n, \mu_\sigma \}$.  We then set $\gamma$ equal to the real part of the eigenvalue with the largest real part.  If there is only one eigenvalue whose real part equals $\gamma$ then in the long-time limit $R(\omega)$ goes to a steady state, the linear model maps onto the Kuramoto model, and there is full phase locking and synchronization (as defined below).  Otherwise, $R(\omega)$ will never go to a steady-state value, and the phases of $\psi$ will never converge.  

\paragraph{Comments} The model (\ref{kurmap}) with coupling constants (\ref{coup}) can be seen as a {\emph{quenched}} limit of the full stochastic Kuramoto model \cite{Sakaguchi}. Though deterministic, our model is more general because it allows for arbitrary coupling constants, while the standard stochastic version is understood only for specific distributions of these constants, with strong monotonicity constraints. As we shall see, the freedom introduced by our model leads to interesting qualitative differences for the behavior of the system in the vicinity of the critical point. 

\section{Analysis of the Kuramoto model under the linear mapping}

Unless otherwise specified, we shall hereafter assume that we are in the situation where  $R(\omega)$ goes to a steady state, and thus eq. (\ref{kurmap}) becomes the Kuramoto model with time-independent coupling.  

We now define our use of the term ``synchronization''.  We restrict the analysis in this paper to systems with no partial population of drifting oscillators, i.e. the incoherent-to-partially locked (usually referred to as the synchronization transition) and the partially locked-to-fully locked phase transitions occur at the same point \cite{locking}.  We say our system is synchronized if the synchronization order parameter given by
\begin{equation}
r=\left|\int_{-\infty}^{\infty} d\omega g(\omega) e^{i \theta(\omega)}\right|=\left|\int_{-\infty}^{\infty} d\omega g(\omega) \frac{\psi(\omega,t)}{|\psi(\omega,t)|}\right|
\end{equation}
goes to a nonzero steady-state value.  As mentioned above, this happens at the critical point where the real part of more than one eigenvalue becomes equal to $\gamma$.  So in the synchronized region where the real part of only one eigenvalue equals $\gamma$,
\begin{equation}
\label{rb}
r=\left|\int_{-\infty}^{\infty} d\omega g(\omega) \frac{b(\omega)}{|b(\omega)|}\right|
\end{equation}
where $b(\omega)$ is the eigenfunction corresponding to that differentiated eigenvalue, $\lambda_N$.

Let us consider now a specific example of global coupling in the linear model, $\Omega(\omega,\omega')=\Omega$ \cite{coup}.  The linear model describes this system as 
\begin{equation}
\label{lo}
\dot \psi(\omega,t)=(i \omega - \gamma) \psi(\omega,t) +\Omega \int_{-\infty}^{\infty} g( \omega') \psi(\omega',t) d \omega',
\end{equation}
which maps onto the original Kuramoto model in the steady state with the following coupling constant:
\begin{equation} \la{koup}
K(\omega,\omega')= \Omega \sqrt{\frac{(\omega - \omega_r)^2 + \gamma^2}{(\omega' - \omega_r)^2 + \gamma^2}},
\end{equation}
where $\omega_r$ is the collective frequency of the synchronized state and is given by the imaginary part of $\lambda_N$, $\Im[\lambda_N]$.   In the following analysis (unless otherwise specified), we will assume that collective frequency $\Im[\lambda_N]$ is zero as it can be eliminated through a phase change in $\psi$,  which effectively puts our system in a rotating reference frame and does not affect the analysis.

With this coupling scheme, we can solve the spectrum of the RHS of eq. (\ref{lo}) exactly.  Fortuitously, eq. (\ref{lo}) maps onto the problem of stability of the fundamental mode of the incoherent state in the Kuramoto model with global uniform coupling \cite{strogatz}, and we can follow the same steps towards a solution.  With $\gamma$ set as described above, the spectrum comprises a continuous line of eigenvalues $\lambda$ in the complex plane along $\Re[\lambda]=-\gamma$ (where $\Re[\lambda]$ denotes the real part of $\lambda$) for any value of $\Omega$ and one eigenvalue $\lambda_N$ at the origin, which stands apart from the continuum of eigenvalues if $\Omega > \Omega_c$.  As $\Omega \rightarrow \Omega_c^+$, $\gamma \rightarrow 0$.  By contrast, for $\Omega \le \Omega_c$ the continuum of eigenvalues lies along the imaginary axis and $\lambda_N$ becomes indistinguishable from the continuum, as shown in Figure 1; and since, in the steady state, the entire spectrum remains, it is clear that $r=0$ where $\Omega \le \Omega_c$.  That (in the steady state of the system) $r \to 0$ as $\Omega \to \Omega_c^+$ indicates that at that point the system is in a splay state \cite{splay}, where the phases of the oscillators in the system are locked in a uniform distribution from 0 to $2\pi$.  Also, above $\Omega_c$, as the coupling between oscillators increases, $r$ grows continuously from $r=0$, a behavior akin to a second-order phase transition.  This will be shown below.

\begin{figure}
\centering
\resizebox{0.75\columnwidth}{!}{
\includegraphics{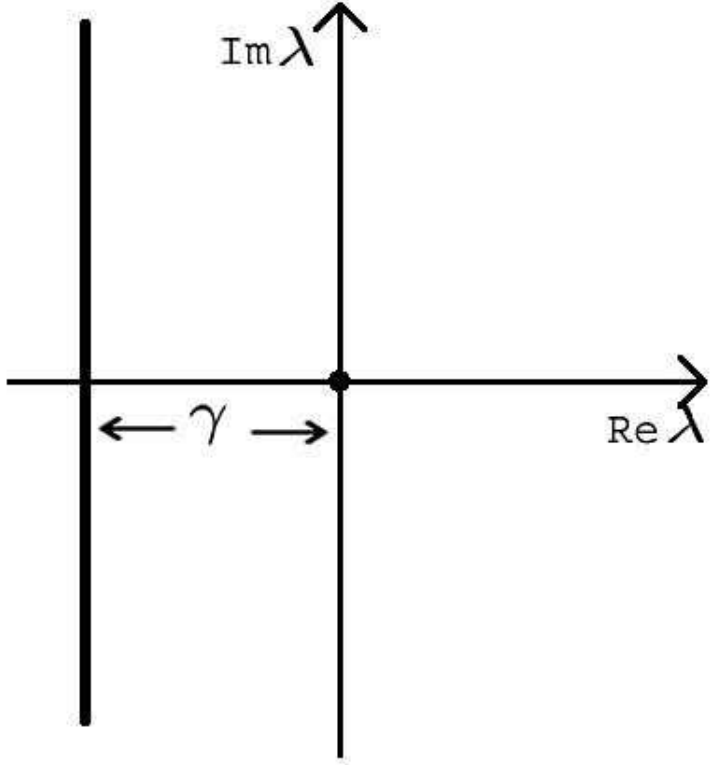} 
}
\caption{The spectrum of eigenvalues associated with the RHS of eq. (\ref{lo}) when $\Omega > \Omega_c$.  The spectrum comprises a continuum of eigenvalues along $-\gamma$ and a single eigenvalue at the origin.  As $\Omega \rightarrow \Omega_c^+$, $\gamma \rightarrow 0$, and the continuum approaches the imaginary axis and the eigenvalue at the origin.  When $\Omega \le \Omega_c$, $\gamma = 0$, the continuum of eigenvalues sits on the imaginary axis and the eigenvalue at the origin becomes indistinguishable from the continuum.}
\label{fig:1}       
\end{figure} 

One can derive the nontrivial behavior of $r$ from eq. (\ref{lo}).  Assuming $\psi(\omega,t)=b(\omega) e^{\lambda_N t}$, eq. (\ref{lo}) becomes
\begin{equation}
\label{lo_eg}
\lambda_N b(\omega)=i \omega b(\omega)+\Omega A
\end{equation}
where $A \equiv \int_{-\infty}^{\infty} g( \omega) b(\omega) d \omega$, which is a constant that can be determined self-consistently.  Treating $A$ as a constant, one can algebraically solve eq. (\ref{lo_eg}) for $b(\omega)$ and then plug this solution for $b(\omega)$ back into the definition of $A$ to arrive at the following formula that determines $\gamma$:
\begin{equation}
\label{sc}
1=\Omega \int_{-\infty}^{\infty} \frac{g( \omega)}{\lambda_N-i \omega} d \omega
\end{equation}
As before, we set $\gamma = \Re[\lambda_N]$. Assuming $g(\omega)$ is an even function and nowhere increasing for $\omega \ge 0$, then eq. (\ref{sc}) has at most one solution for $\gamma$ and $\gamma>0$ \cite{mirollo}, which implies
\begin{equation}
\label{Oc}
1=\Omega \int_{-\infty}^{\infty} \frac{g( \omega) \gamma}{\gamma^2+ \omega^2} d \omega
\end{equation}
By taking $\gamma \rightarrow 0^+$ ($r \to 0$ as discussed above) and assuming $g(\omega)$ has a finite width, it becomes clear from eq. (\ref{Oc}) that
\begin{equation}
\label{crit}
\Omega_c=\frac{1}{\pi g(0)}.
\end{equation}
Knowing that $b(\omega)=\frac{\Omega A}{\lambda_N-i \omega}$ from eq. (\ref{lo_eg}), for $\Omega > \Omega_c$ (i.e. $r>0$) we can also compute an explicit expression for $r$ from eq. (\ref{rb}) (recalling that $g(\omega)$ is even) to arrive at
\begin{equation}
\label{rgen}
r = \int_{-\infty}^\infty d\omega g(\omega) \frac{1}{\sqrt{1+\left(\frac{\omega}{\gamma}\right)^2}}
\end{equation}
where $\gamma$ can be determined from eq. (\ref{Oc}).  So, for a given distribution $g(\omega)$, eqns (\ref{Oc}), (\ref{crit}), and (\ref{rgen}) completely specify $r(\Omega)$ and $\Omega_c$ for $K(\omega,\omega')$, eq. (\ref{koup}).

\subsection{Scaling properties of the order parameter and coupling constant near the critical point}

Equations (\ref{Oc}) and (\ref{rgen}) give interesting, new behavior (resembling a second-order phase transition) in the critical regime. In order to study eq. (\ref{Oc}) perturbatively, we start from its form (\ref{sc}) and the identity $\p_{\lambda_N + i \omega}f(\lambda_N - i\omega) = 0$, which is valid for any smooth function $f$. Denoting (in a distribution sense, assuming that we work in the correct Sobolev space \cite{distrib})
\be
{\cal D}_{\gamma}(\omega) \equiv \frac{1}{\gamma - i \omega}, 
\ee
we can expand around the origin to obtain
\be
{\cal D}_{\gamma}(\omega) = {\cal D}_{0}(\omega) + \gamma [\p_{\gamma} {\cal D}_{\gamma}(\omega)]_{\gamma = 0} + O(\gamma^2).
\ee
Since ${\cal D}_{0}(\omega) = \pi \delta(\omega)$ ({\sf {c.f.}} (\ref{crit})), using the identity mentioned above, (\ref{Oc}) becomes
\be
1 = \Omega \left[ \pi g(0) - \Re \left( i \gamma \int_{-\infty}^{\infty} \frac{g'(\omega)}{\gamma - i \omega} d \omega \right) \right], \quad {\rm {for} } \quad\gamma \to 0,
\ee
where $g'(\omega)$ is the derivative of $g(\omega)$ with respect to $\omega$.  This can be expressed as
\be
1 = \Omega \left( \pi g(0) + \gamma \lim_{\gamma\to 0}\int_{-\infty}^{\infty} \frac{\omega g'(\omega)}{\gamma^2 + \omega^2} d\omega \right),
\ee
which finally yields
\be
1 = \Omega \left( \pi g(0) + \gamma \int_{-\infty}^{\infty} \frac{g'(\omega)}{\omega} d\omega \right).
\ee
Perturbatively, the behavior of $\gamma$ as $\Omega \rightarrow \Omega_c$ becomes 
\begin{equation}
\la{gammascale}
 \gamma= - \frac{\pi g(0)}{\int_{-\infty}^\infty d\omega g'(\omega)/\omega} \left( \frac{\Omega-\Omega_c}{\Omega_c} \right) + O\left[ \left(\frac{\Omega-\Omega_c}{\Omega_c} \right) ^2 \right],
\end{equation}
where we have assumed the distribution $g(\omega)$ is such that $\int_{-\infty}^\infty d\omega g'(\omega)/\omega$ is nonzero and finite.

Let us now turn to eq. (\ref{rgen}). Using the Weber formula for the Laplace transform of the Bessel function of first kind,
\be
\frac{1}{\sqrt{\omega^2 + \gamma^2}} = \int_0^{\infty} e^{-\omega s} J_0(\gamma s) ds,
\ee
we obtain 
\be \la{hankel}
r = \gamma \int_0^{\infty} \tilde{g} (s) J_0(\gamma s) ds,
\ee
where $\tilde{g}(s) = m(-s)$ is the Laplace transform of $g(x)$, or the moment-generating function of the negative argument $-s$. Alternatively, we may say that $r$ is proportional to the Hankel transform of 
$\tilde{g}(s)/s$. Eq. (\ref{hankel}) allows us to extract the behavior of $r(\gamma)$ as $\gamma \to 0$. 
First, notice that, since we are interested in the limit of vanishing $\gamma$, we may consider the test p.d.f.'s with compact support $g_L(\omega) = {\sf {U}}[0,L]$ from the uniform family. Then the Laplace transform becomes
\be \la{laplace}
\tilde{g}_L(s) = \frac{1-e^{-Ls}}{Ls} \equiv \nu(y), \quad y = Ls,
\ee
where we introduce the scaled variable $y = Ls$. Clearly, the scaled function $\nu$ has the behavior
\be \la{beh}
\begin{array}{ll}
\nu (y) \simeq \frac{1}{y}, & \quad {\mbox {for}} \quad y \gg 1, \\
& \\
\nu (y) \to 1, & \quad {\mbox {for}} \quad y \ll 1, 
\end{array}
\ee
so that eq. (\ref{hankel}) becomes
\be \la{int}
r(\gamma) = \frac{2 \gamma}{L} \int_0^\infty \nu(y) J_0(\nu y/L) d y,
\ee
where the factor of 2 arises because we take the uniform distribution from $-L$ to $L$.  Let $x_*$ be the smallest zero of $J_0, J_0(x_*) = 0$. The integral eq. (\ref{int}) may then be approximated 
as 
\be
r(\gamma) = \frac{2 \gamma}{L}\int_\epsilon^{\frac{L}{\gamma}x_* } \frac{dy}{y} + O \left ( 
\frac{\gamma}{L} \right ),
\ee
where $\epsilon$ is an {\emph {infrared}} cut-off reflecting the small-$y$ behavior of $\nu$ eq. (\ref{beh}). 
Asymptotically in the small parameter $\gamma/L$, this gives the expansion
\be \la{log}
r(\gamma) = - 2 g(0) \gamma \log[g(0) \gamma] + O[g(0) \gamma], \quad  \gamma \to 0,
\ee
where $g(0) = \frac{1}{L}$.  It is also worth mentioning that $r(\gamma)$ is a monotonically increasing function taking values from 
$r(0) = 0$ to $r(\gamma \to \infty) = 1$, so it is a properly defined {\emph {cumulative distribution function}} on $\mathbb{R}_+$.

\subsection{Specific examples of distributions}

With these general solutions for the parameters of any system with global coupling $\Omega(\omega,\omega')=\Omega$, we can solve for a specific system given its characteristic frequency distribution.  Take for instance the Lorentzian distribution about $\omega_r$, i.e. $g(\omega-\omega_r) = \frac{\Delta}{\pi [\Delta^2+(\omega-\omega_r)^2]}$.  From eq. (\ref{crit}) we find $\Omega_c=\Delta$, and from eq. (\ref{sc}), $\gamma_{lor} = \Omega-\Delta$.  Using these in eq. (\ref{rgen}), we obtain
\begin{equation}
r_{lor}=\frac{2 \cos^{-1} \left( \frac{\Omega_c}{\Omega-\Omega_c} \right)}{\pi \sqrt{1-\left( \frac{\Omega_c}{\Omega-\Omega_c} \right)^2}}
\end{equation}
for $\Omega > \Omega_c$, as shown in Figure 2. $r=0$ for $\Omega \le \Omega_c$ as discussed above.  From a Taylor expansion around the origin of this expression, we obtain the following scaling for this distribution:
\begin{equation}
r_{lor} \approx \frac{2}{\pi} \frac{\Omega-\Omega_c}{ \Omega_c} \log \left( \frac{\Omega_c}{\Omega-\Omega_c} \right).
\end{equation}
This agrees with eqns (\ref{gammascale}) and (\ref{log}) knowing that, for this Lorentzian distribution, $g(0) = \frac{1}{\pi \Delta}$ and $\int_{-\infty}^\infty d\omega g'(\omega)/\omega = - \frac{1}{\Delta^2}$.

Similarly, for a uniform distribution about $\omega_r$, i.e. $g(\omega-\omega_r) = \frac{1}{\pi \Delta}$ for $|\omega-\omega_r| < \pi \Delta/2$ and 0 otherwise, the above equations give $\Omega_c=\Delta$, $\gamma_{unif}=\frac{\Delta\pi}{2} \cot\left(\frac{\pi \Delta}{2 \Omega}\right)$, and
\begin{equation}
r_{unif}=\cot \left( \frac{\pi}{2} \frac{\Omega_c}{ \Omega} \right) \sinh^{-1} \left[ \tan \left( \frac{\pi}{2} \frac{\Omega_c}{\Omega  }\right)\right]
\end{equation}
for $\Omega > \Omega_c$ (see Figure 2).  As above, $r=0$ for $\Omega \le \Omega_c$.  Using a Taylor expansion about the origin, the scaling for the uniform distribution is then
\begin{equation}
r_{unif} \approx \frac{\pi}{2} \frac{\Omega-\Omega_c}{ \Omega_c}\log \left( \frac{\Omega_c}{\Omega-\Omega_c} \right).
\end{equation}
Again, there is agreement with eqns (\ref{gammascale}) and (\ref{log}) as $g(0)=\frac{1}{\pi \Delta}$ and $\int_{-\infty}^\infty d\omega g'(\omega)/\omega = - \frac{4}{(\pi \Delta)^2}$.

\begin{figure}
\centering
\resizebox{.95 \columnwidth}{!}{
\includegraphics{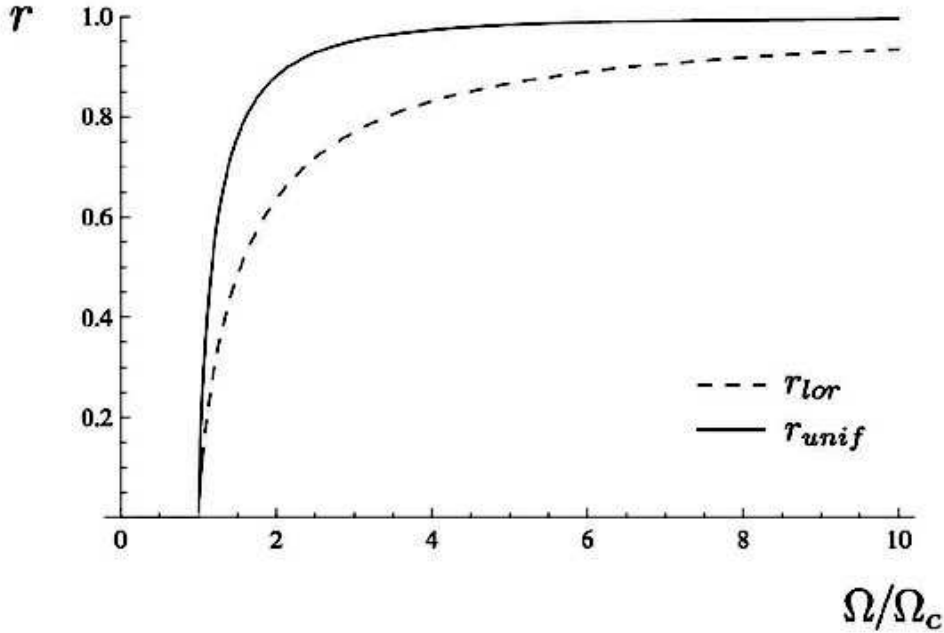} 
}
\caption{The synchronization order parameter as a function of the normalized coupling constant for a uniform and a Lorentz distribution of characteristic frequencies.}
\label{fig:2}       
\end{figure}

\section{Discussion}

The synchronization phase transition discussed in this paper is of the {\em nonequilibrium} variety, but there are many interesting connections with the more typical equilibrium phase transitions.  One way to illuminate the nature of this nonequilibrium phase transition and create a platform for comparison with standard theories of equilibrium phase transitions would be to determine the critical behavior of the fluctuations of the order parameter close to the critical point, which exhibit power-law scaling in the uniform Kuramoto model \cite{daido}.

Although the physical reason for the logarithmic scaling behavior of the synchronization order parameter close to the critical point in the linear model eq. (\ref{model}) remains unclear, we wish to highlight a number of distinctive features of this phase transition that may be a cause of such behavior:
\begin{itemize}
\item[(1)]  The specific coupling considered in this paper (eq. 10) is of a very different nature from the standard uniform coupling scheme associated with the Kuramoto model.  It is therefore unsurprising that the scaling behavior of the linear reformulation also differs from the usual square-root scaling behavior of the normal Kuramoto model close to the critical point. This particular coupling is asymmetric and has the interesting property that the coupling between two oscillators becomes stronger the closer one oscillator's characteristic frequency is to $\omega_r$ and the further the other's characteristic frequency is from $\omega_r$ (reminiscent of problems involving asymptotic freedom \cite{freedom}).  It should also be noted that close to the critical point, i.e. $\gamma \to 0$, if the characteristic frequency of one of the oscillators is $\omega_r$ then the coupling tends to infinity in one direction and zero in the other. \\
\item [(2)] The standard Kuramoto model describes two phase transitions | one from incoherence to partial locking and a second from partial locking to full locking | which in general occur at different points in parameter space \cite{locking}.  In our paper we consider systems where these two phase transitions are coincident, i.e. where a bicritical point occurs. \\
\item[(3)] This logarithmic scaling near a critical point is interestingly reminiscent of the behavior of the marginal eigenvalues in standard renormalization group theory \cite{goldenfeld}.
\end{itemize}

Although this paper only explored the critical properties of one particular type of coupling, the full range of coupling schemes $K(\omega,\omega')$ (eq. (\ref{coup})) that can be analyzed with this method remains to be determined.

Acknowledgements:  D.C.R. gratefully acknowledges stimulating discussions with Steven Strogatz and Matthew Hastings.  R.T. acknowledges support from the LDRD Directed Research grant on Physics of Algorithms.  This research was carried out under the auspices of the National Nuclear Security Administration of the U.S. Department of Energy at Los Alamos National Laboratory under Contract No. DE AC52-06NA25396.


\begin{thebibliography}{1}
\bibitem{coup}  The following analysis is also valid for more general couplings of the form $\Omega(\omega,\omega')=\alpha(\omega) \beta(\omega')$.
\end{thebibliography}
\end{document}